\documentclass[twocolumn,aps, prl, superscriptaddress, showpacs, amsmath,amssymb]{revtex4}
\usepackage{graphicx}
\usepackage{dcolumn}
\usepackage{bm}
\begin{document}
\title{Collective behavior in a granular jet: Emergence of a liquid with zero surface-tension}
\author{Xiang Cheng}
\affiliation{The James Franck
Institute and Department of Physics, The University of Chicago,
Chicago, IL 60637}
\author{German Varas}
\altaffiliation [Present address: ] {Departmento de Fisica,
Universidad de Chile, Casilla 487-3, Santiago, Chile.}
\affiliation{The James Franck Institute and Department of Physics,
The University of Chicago, Chicago, IL 60637}
\author{Daniel Citron}
\affiliation{The James Franck Institute and Department of Physics,
The University of Chicago, Chicago, IL 60637}
\author{Heinrich M. Jaeger}
\affiliation{The James Franck Institute and Department of Physics,
The University of Chicago, Chicago, IL 60637}
\author{Sidney R. Nagel}
\affiliation{The James Franck Institute and Department of Physics,
The University of Chicago, Chicago, IL 60637} 
\date{\today}

\begin{abstract}
We perform the analog to the ``water bell'' experiment using
non-cohesive granular material.  When a jet of granular material,
many particles wide, rebounds from a fixed cylindrical target, it
deforms into a sharply-defined sheet or cone with a shape that
mimics a liquid with zero surface tension.  The particulate nature
of granular material becomes apparent when the number of particles
in the cross-section of the jet is decreased and the emerging sheets
and cones broaden and gradually disintegrate into a broad spray.
This experiment has its counterpart in the behavior of the
quark-gluon plasma generated by collisions of gold ions at the
Relativistic Heavy Ion Collider. There a high density of
inter-particle collisions gives rise to collective behavior that has
also been described as a liquid.
\end{abstract}

\maketitle When one or two particles strike a smooth wall at normal
incidence, they rebound in the direction whence they came. Yet, as
we show here, a dense stream of non-cohesive particles hitting a
target retains its integrity and deforms into a thin sheet with a
shape resembling the structures created by an impinging water jet
\cite{savart,clanet}. Thus the collective behavior of many particles
differs qualitatively from that of the individual components. When
can a jet of discrete particles be modeled as a liquid
\cite{review1,review2,review3,Pouliquen} and how do the liquid
patterns emerge out of individual particle scattering events from a
target? Such questions, posed and investigated here with granular
materials, have their counterpart in much more microscopic
situations such as the quark-gluon plasma caused by relativistic
high-energy collisions of gold ions, which also produces scattering
patterns indicative of a liquid state \citet{RHICBrook,RHICori}. Our
findings provide a macroscopic, purely classical example of how
strong interactions, mediated by rapid collisions in a densely
packed region, can give rise to liquid behavior.

When a water stream hits a flat target, it spreads symmetrically in
the direction transverse to the impact and deforms into a thin
sheet.  For targets smaller than the stream diameter, the sheet
forms a hollow bell-shaped structure that envelopes the target. Such
"water bells" were reported in 1883 by Savart \cite{savart} and have
been systematically studied by Clanet \cite{clanet}.  By contrast,
with the exception of several simulations and experiments on
granular streams passing by solid obstacles at relatively low
particle density and speed \cite{poschel,kellay,swinney}, little is
known about the structures emerging from equivalent experiments
using granular material rather than a liquid. Here, we examine a
non-cohesive granular jet colliding with a target. For jets many
particles wide, the impact produces granular sheets and cones
similar to those seen for water with structures that depend on the
ratio of the jet to the target diameter. In this regime, we conclude
that the granular medium behaves as if it were a liquid with
infinite Weber number -- that is, with zero surface tension
appropriate for a system of non-cohesive particles.  The particulate
nature of the material becomes apparent when the sheets and cones
broaden and gradually disintegrate as the number of particles in the
beam is decreased.

\begin{figure}
\begin{center}
\includegraphics[width=3.3in]{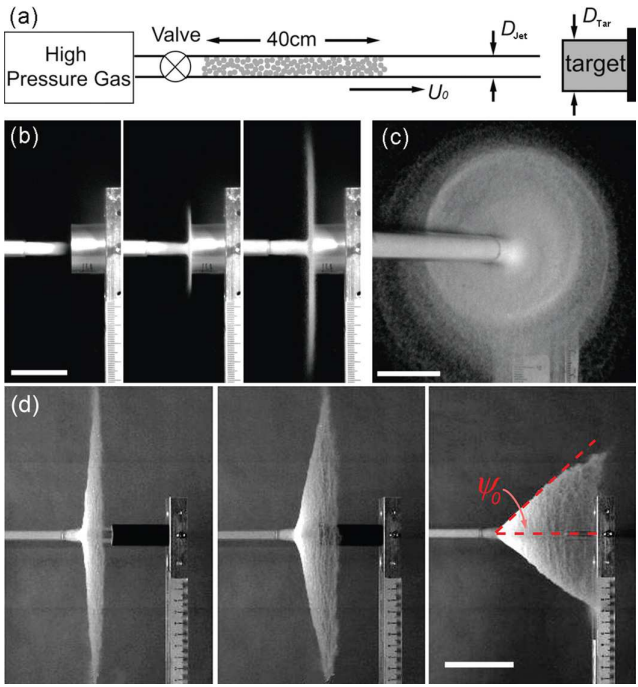}%
\caption{\label{Figure1} (color online). Liquid-like granular sheets
and cones. (a) Sketch of the experimental setup. (b) Side view of a
granular jet comprised of 100$\mu$m glass beads hitting a target at
velocity $U_0\sim10$m/s.  The ratio of target to jet diameter is
$D_{Tar}/D_{Jet}=3.80$. The images show the jet 0.5ms before as well
as 2.5ms and 9.5ms after impact (left to right). Scale bar is 3.0cm.
(c) Axial view of the same granular sheet 35.5ms after impact. Scale
bar is 2.5cm. (d) Side views of the hollow cones produced by
100$\mu$m glass bead jets for smaller target-to-jet diameter ratios.
$D_{Tar}/D_{Jet}=2.0$, 1.62, and 0.88 (left to right). Scale bar is
6.0cm.}
\end{center}
\end{figure}

\begin{figure}
\begin{center}
\includegraphics[width=3.2in]{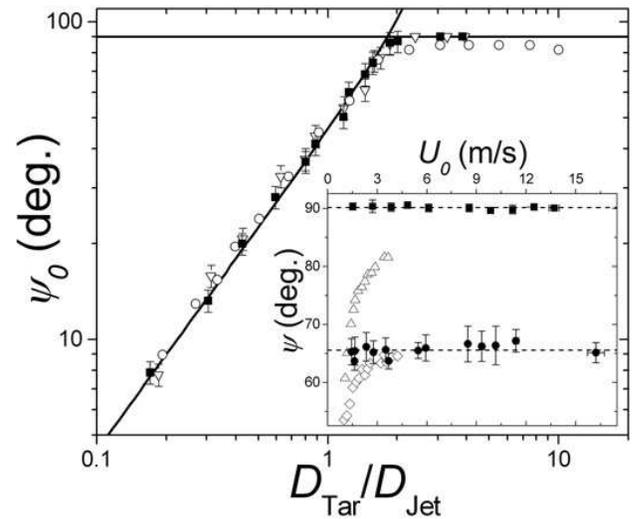}%
\caption{\label{Figure2} Opening angle of the granular sheets and
cones. Angle $\psi_0$ as a function of $D_{Tar}/D_{Jet}$ for
100$\mu$m glass beads ($\blacksquare$), 100$\mu$m copper beads
($\triangledown$) and water in the large-We limit ($\circ$). The
only observable difference is that $\psi_0$ at large
$D_{Tar}/D_{Jet}$ is $\sim85^\circ$ for water but $90^\circ$ for
grains. The solid lines are fits of Eq.\ref{eq1} with $(A-B)=0.30$
for $D_{Tar}/D_{Jet}<2$. Inset: Angle $\psi$ as a function of jet
velocity $U_0$ for 100$\mu$m glass beads with $D_{Tar}/D_{Jet}=2.47$
($\blacksquare$) and $D_{Tar}/D_{Jet}=1.32$ ($\bullet$), and for
water with $D_{Tar}/D_{Jet}=3.29$ ($\vartriangle$) and
$D_{Tar}/D_{Jet}=1.33$ ($\Diamond$). The horizontal lines are
$\psi=90^\circ$ (top) and $\psi=65.5^\circ$ (bottom). Data for water
is taken from \cite{clanet}.}
\end{center}
\end{figure}

We prepared our dense jets by packing granular material into a 40cm
section of a glass launching tube of inner diameter $D_{Jet}=0.73$cm
(Fig.\ref{Figure1}a).  The grains, which were compacted to a
reproducible density by tapping, were mono-disperse spherical beads
of glass ($\rho=2.5$g/cm$^3$) or copper ($\rho=8.2$g/cm$^3$) with
diameters between $d=50\mu$m and $d=2.1$mm. Prior to filling the
tube, we baked the beads in vacuum to minimize any residual adhesion
between beads.  Pressurized gas accelerated this granular plug into
a jet which hits a target 2.5cm in front of, and collinear with, the
tube. The jet velocity $U_0$ could be varied between 1 and 16m/s.
The impacts were filmed at 2000 frames/second with a Phantom V7.1
camera. Fig.\ref{Figure1}b and c show the side and front views
respectively of a colliding jet. Since there are very few collisions
inside it, the jet maintains its cylindrical shape before hitting
the target, consistent with jets produced when a sphere impacts
loosely packed powers \cite{lohse,royer}. After hitting the target,
the jet deforms into an extraordinarily thin symmetric granular
sheet clearly resembling a spreading liquid. To investigate the
similarities between granular and ordinary liquid jets, we first
keep the jet diameter, $D_{Jet}$, fixed at values much larger than
the particle diameter, $d$, and vary the target diameter, $D_{Tar}$,
as shown in Fig.\ref{Figure1}d. When $D_{Tar}$ is reduced, these
planar sheets change into cones with opening angle,
$\psi_0<90^\circ$. Fig.\ref{Figure2} shows that $\psi_0$ increases
linearly with $D_{Tar}/D_{Jet}$ until $\psi_0$ saturates at
$90^\circ$ above $D_{Tar}/D_{Jet}\sim2$. Glass and copper beads
produces essentially identical behavior.

We can compare our results with those from Clanet \cite{clanet} for
water jets, who found that the opening angle, $\psi$, of the ``water
bell'' depends on the Weber number $\textrm{We}=\rho
U_0^2D_{Jet}/\sigma$, where $U_0$ is the jet velocity and $\rho$ the
density and $\sigma$ the surface tension of water. In the large-We
limit (large $U_0$ or small $\sigma$), $\psi$ approaches a constant,
$\psi_0$,  that depends only on $D_{Tar}/D_{Jet}$. Fig.\ref{Figure2}
shows excellent overlap of our values for $\psi_0$ with those of
Clanet. As shown in the inset to Fig.\ref{Figure2}, $\psi$ for
granular material remains constant as $U_0$ is varied over our
entire experimental range (an order of magnitude) whereas $\psi$ for
water decreases at small velocities. This suggests a granular jet
impacting a the target behaves like a liquid with negligible
surface-tension: regardless of the value of $U_0$, the Weber number
is pinned at infinity since the absence of significant cohesive
forces effectively drives the surface tension to zero.

The linear dependence of $\psi_0$ on $D_{Tar}/D_{Jet}$ in the
large-We limit can be understood from momentum conservation. The
magnitude of the momentum of the incoming granular jet reaching the
target in time $\tau$ is $P_{in}=(\pi/4)D_{Jet}^2U_0^2\tau$. The
collisions of the jet with the target are inelastic so the outgoing
momentum magnitude $P_{out}=CP_{in}$ where $C$ is the average
coefficient of restitution. Along the axis, the momentum balance is
given by:
$P_{in}-P_{out}\cos\psi_0=P_{in}(1-C\cos\psi_0)=F_{Tar}\tau$, where
$F_{Tar}$ is the average force along the axis exerted by the target
on the jet. Transverse to the axis, the momentum (averaged over the
entire sheet) remains zero. When $D_{Tar}/D_{Jet}<1$,
$F_{Tar}=(A\pi/4)D_{Tar}^2\rho U_0^2$. Here $A$ is a constant
describing the average glancing collision angle for a particle.  In
this region, the average restitution coefficient, $C$, also depends
on the fraction of particles hitting the target: $C=[1
-B(D_{Tar}/D_{Jet})^2]$ where $B$ depends on the coefficient of
restitution for single collisions. When $D_{Tar}/D_{Jet}\gg1$, the
entire momentum of the jet is reflected by the target.  Finally we
obtain:
\begin{equation}
\psi_0=\left\{\begin{array}{ll}
\textrm{acos}(1-(A-B)(\frac{D_{Tar}}{D_{Jet}})^2); \quad \frac{D_{Tar}}{D_{Jet}}\ll1\\
 \\
90^\circ; \quad \frac{D_{Tar}}{D_{Jet}}\gg1
\end{array} \right.  \label{eq1}
\end{equation}
Clanet reached the same result by considering the momentum transfer
during the impact of a water jet on a target using hydrodynamic
equations \cite{clanet}. By fitting the experimental data in
Fig.\ref{Figure2} we find $(A-B)=0.30\pm0.02$ which is close to the
value $\approx0.352$ for water \cite{clanet}.

\begin{figure}
\begin{center}
\includegraphics[width=3.35in]{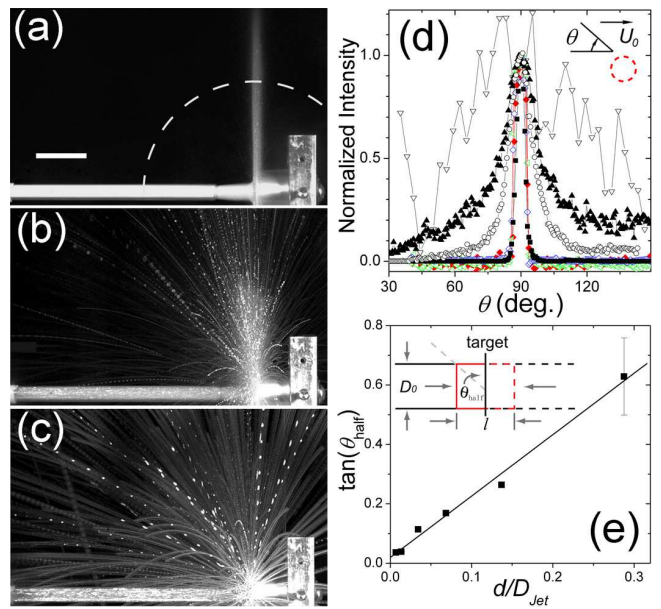}%
\caption{\label{Figure3} (color online). Scattering patterns as a
function of particle density across the jet diameter. (a)-(c)
Maximum intensity projections of the scattering process for fixed
$D_{Tar}/D_{Jet}=3.8$ but varying particle density: (a)
$D_{Jet}/d=73$; (b) $D_{Jet}/d=14.6$; (c) $D_{Jet}/d=3.5$. (d)
Normalized scattering profiles for glass beads in air with
$D_{Jet}/d=73$ ($\blacksquare$), $D_{Jet}/d=29.2$ ($\circ$),
$D_{Jet}/d=14.6$ ($\blacktriangle$), $D_{Jet}/d=3.5$
($\triangledown$); glass beads in helium $D_{Jet}/d=73$ (red
$\blacklozenge$); copper beads in air $D_{Jet}/d=73$ (green
$\vartriangleleft$) and rough sand in air $D_{Jet}/d=73$ (blue
$\lozenge$). The arrow and the dashed circle in the inset indicate
the direction and the cross-section of the jet. (e) Half-width
$\theta_\textrm{half}$ of the angular scattering profiles in (d) as
a function of $d/D_{Jet}$. Solid line is a linear fitting. The model
is sketched in the inset and explained in the text.}
\end{center}
\end{figure}

The impact of dense particle streams clearly generates similar
patterns as do liquids. How does the particulate nature of granular
material becomes manifest as the number of particles within the jet
decreases?  To vary this number we change the ratio, $D_{Jet}/d$, of
jet to particle diameter. In the limit $D_{Tar}/D_{Jet}\gg1$,
Figs.\ref{Figure3}a-c show major qualitative changes in the particle
trajectories as $D_{Jet}/d$ is reduced. The images were created by
superimposition of many different consecutive still images; each
pixel shows the maximum intensity at that location over all images
in the time period. For $D_{Jet}/d=73$, almost all particles emerge
in a sheet normal to the jet axis. For $D_{Jet}/d=14.6$, the sheet
becomes more diffuse as some particles leave the plane as shown by
the bright lines. For $D_{Jet}/d=3.5$, a firework-like pattern
results after impact and sheet structure is no longer apparent as
particles rebound from the target in a broad angular distribution.
To quantify this trend, we plot in Fig.\ref{Figure3}d the angular
scattering distribution, obtained by averaging images as in
Fig.\ref{Figure3}a-c along a circle of radius $r=8.7D_{Jet}$
centered on the jet axis (shown partially in Fig.\ref{Figure3}(a) as
the dashed line). The axial position of the center is chosen to be
in the middle of the ejected particles, which gradually moves away
from the target in upstream direction as the particle diameter $d$
increases. To avoid the target holder, only the scattering profile
along the upper half of the circle is plotted. Each profile is
normalized to its peak value. The scattering angle $\theta$ is zero
along the axis of the jet and increases clockwise. As $D_{Jet}/d$
decreases, these profiles become broader.

To investigate this crossover to diffuse scattering we examined the
roles of air and inelasticity during inter-particle collisions. To
exclude air as the primary cause of the sheet formation, three
experiments were performed. (i) Doing the entire experiment in a
helium atmosphere and using helium as the accelerating gas produced
no change in the angular distribution, as shown in
Fig.\ref{Figure3}d for $D_{Jet}/d=73$. (ii) Performing our
experiment at reduced ambient pressure of 31Pa also resulted in a
thin, planar sheet structure. (31Pa corresponds to a mean free path
of air molecules three times the 100$\mu$m grain diameter used in
Fig.\ref{Figure3}a. Although air accelerates the granular column
from behind, we estimate that the velocity of air penetrating into a
40cm granular pack comprised of 100$\mu$m beads with packing
fraction around 0.6 is two orders of magnitude smaller than the
velocity of the jet. Thus, by the time the front of the jet hits the
target, a negligible amount of air has entered the chamber.) (iii)
We decreased the tube diameter $D_{Jet}$, but fixed the particle-air
interaction by keeping $d=100\mu$m. In this case the planar
structure shown in Fig.\ref{Figure3}a disappeared gradually, and for
$D_{Jet}\sim d$ we regained the firework pattern shown in
Fig.\ref{Figure3}c. In addition, we altered the inter-particle
collision dynamics by changing the particle material or the surface
roughness. Fig.\ref{Figure3}d shows that for copper particles that
are less elastic than glass or for rough sand particles the angular
scattering distribution remains unchanged. However, distributions
for the same $D_{Jet}/d$ do not overlap quantitatively for different
$d$, and the liquid-like sheets tend to be more sharply delineated
with smaller particles.

While we cannot rule out that inelastic collisions or effects of air
affect the detailed shape of the scattering profiles, our results
imply that the creation of the sheets arises fundamentally from the
rapid collisions occurring in an interaction region right in front
of the target.  We found similar behavior when two granular jets
collide head-on, implying that the target serves primarily to
reverse the direction of particles incident upon it. For this
situation, a simple geometric model can capture the essence of the
crossover from fluid- to granular behavior.  As shown in
Fig.\ref{Figure3}e, inset, we divide the system into three zones:
two external regions in which the jets are traveling towards one
another but have not yet collided, and an interaction region with
the same diameter as the jets, $D_{Jet}$, and axial length, $l$.
Inside the interaction region particles undergo rapid collisions and
are confined by pressure from the incoming jets on both sides. The
only way in which particles can escape (which they must since new
particles are entering the region continuously) is to emerge
perpendicularly to the jet axis. A measure for deviations from this
transverse axis and thus for the half-width of the angular
distributions in Fig.\ref{Figure3}d is the angle
$\theta_\textrm{half}$ by which particles can escape ballistically
from the center of the beam.  By geometry, this is given by
$\tan(\theta_\textrm{half})=l/(D_{Jet})$. For dense particle streams
where the mean free path is much smaller than the diameter of the
particles, we expect $l\sim2d$ since that is the smallest region
occupied by the two colliding particles. Therefore, the half-width
should scale as $\theta_\textrm{half}=\arctan(A2d/D_{Jet})$ where
$A$ is a constant O(1). This dependence on particle density
$d/D_{Jet}$ across the beam is indeed born out by the data in
Fig.\ref{Figure3}e, with $A=1.05\pm0.05$. As a result, when
$D_{Jet}/d$ is sufficiently large the majority of particles undergo
rapid, multiple collisions in front of the target and are ejected
into a narrow angular, similar to a liquid film. For example,
$\theta_\textrm{half}<2^\circ$ for jets with $D_{Jet}/d>60$.

\begin{figure}
\begin{center}
\includegraphics[width=3.35in]{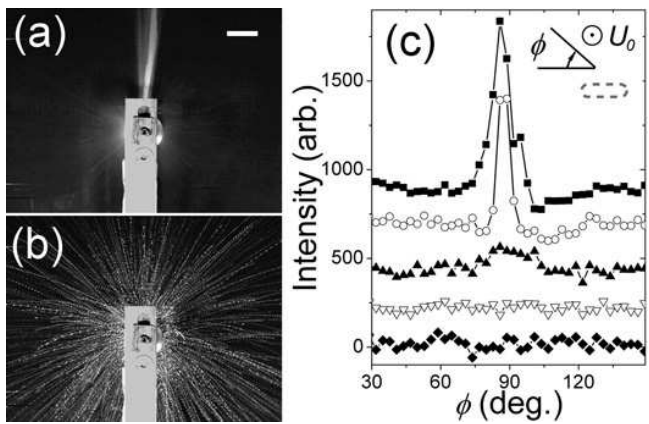}%
\caption{\label{Figure4} Azimuthal scattering profiles for granular
jets with rectangular cross-section (aspect ratio 2).  Shown are
views along the axial direction, looking at the target from behind.
(a) Anisotropic profile resulting for $D_{Jet}/d=73$, where
$D_{Jet}$ is the long axis of the jet cross-section. The target
holder blocks the view of the scattered portion shooting downward.
Scale bar is 3.0cm. (b) Isotropic profile resulting for
$D_{Jet}/d=7$. (c) Azimuthal profiles along a circle with
$r=8.7D_{Jet}$ for $D_{Jet}/d=73$ ($\blacksquare$), $D_{Jet}/d=29.2$
($\circ$), $D_{Jet}/d=14.6$ ($\blacktriangle$), $D_{Jet}/d=7$
($\triangledown$) and $D_{Jet}/d=3.5$ ($\blacklozenge$).The
direction and the cross section of jet are shown in the inset;
$\phi$ is taken to be zero along the long axis of the jet
cross-section.}
\end{center}
\end{figure}

When the granular scattering mimics that of a liquid, we expect
azimuthally anisotropic patterns to be created by jets with
rectangular cross-sections.  The incoming jets establish an
elongated interaction region within the plane transverse to the beam
(azimuthal $\phi$-direction) so that more particles will escape
along the direction of the short axis of the cross-section where a
greater pressure gradient exists. This is also the situation
pertinent to non-central collisions of heavy ions at the
Relativistic Heavy Ion Collider (RHIC) where partial overlap of the
colliding ions establishes an elongated interaction zone for the
quark-gluon plasma.  Azimuthally anisotropic scattering patterns
have been taken as the main evidence for liquid-like behavior of
this plasma \cite{RHICBrook,RHICori}. As shown in Fig.\ref{Figure4}a
for $D_{Jet}/d\gg1$, we observe sharply focused azimuthal patterns
for granular jets with rectangular cross-section. To quantify the
anisotropy and compare with the RHIC results
\cite{RHICBrook,RHICori,RHICtheory}, we analyze the second
coefficient, $\nu_2$, of the Fourier expansion of the azimuthal
scattering profiles in Fig.\ref{Figure4}c. For $D_{Jet}/d=73$ we
obtain $\nu_2=0.16$ which is larger than the typical value at the
RHIC experiments, indicating a stronger interaction between
component particles.  As Figs.\ref{Figure4}b and c show, the
scattering patterns become more isotropic with decreasing
$D_{Jet}/d$.

These experiments demonstrate how non-cohesive granular material can
produce collective motion reminiscent of liquids. Our results
suggest that rapid particle collisions in a very narrow interaction
region, with thickness of a few particle diameters, eject very thin
granular sheets that mimic the liquids without surface tension.  The
crossover between diffuse and sharp scattering profiles appears to
be controlled primarily by the number of collisions in the jet.
These findings may have analogs in other, disparate parts of
physics, where a high density of collisions dominates the behavior.

\begin{acknowledgments}
We thank R. Bellwied, E. Corwin, S. Gavin, T. P\"{o}schel, J. Royer,
T. Witten and L. Xu. This work was supported by the NSF through its
MRSEC program under DMR-0213745 and through its Inter-American
Materials Collaboration under DMR-0303072.

\end{acknowledgments}


\begin{thebibliography}{99}
\bibitem{savart} F. Savart, Ann. de Chim. {\bf 54}, 56 (1833);
Ann. de Chim. {\bf 54} 113 (1833).
\bibitem{clanet} C. Clanet, J. Fluid Mech. {\bf 430}, 111 (2001).
\bibitem{review1} L.P. Kadanoff, Rev. Mod. Phys. {\bf 71}, 435 (1999).
\bibitem{review2} I.S. Aranson and L.S. Tsimring, Rev. Mod. Phys. {\bf 78}, 641 (2006).
\bibitem{review3} H.M. Jaeger, S.R. Nagel, and R.P. Behringer, Rev. Mod. Phys. {\bf 68}, 1259 (1996).
\bibitem{Pouliquen} P. Jop, Y. Forterre, and O. Pouliquen, Nature {\bf 441}, 727 (2006).
\bibitem{RHICBrook} BRAHMS, PHENIX, PHOBOS, and STAR Collaborations. {\it Hunting the
Quark Gluon Plasma: Results from the First 3 Years at RHIC} (Upton,
NY: Brookhaven National Laboratory report No. BNL-73847-2005).
\bibitem{RHICori} I. Arsene {\it et al.} (BRAHMS Collaboration), Nucl. Phys. A {\bf 757}, 1 (2005); B.B. Back
{\it et al.} (PHOBOS Collaboration), Nucl. Phys. A {\bf 757}, 28
(2005); J. Adams {\it et al.} (STAR Collaboration), Nucl. Phys. A
{\bf 757}, 102 (2005); K. Adcox {\it et al.} (PHENIX Collaboration),
Nucl. Phys. A {\bf 757}, 184 (2005).
\bibitem{poschel} V. Buchholtz and T. P\"{o}schel, Granul. Matter {\bf 1}, 33 (1998).
\bibitem{kellay} Y. Amarouchene, J.F. Boudet, and H. Kellay, Phys. Rev. Lett. {\bf 86}, 4286 (2001).
\bibitem{swinney} E.C. Rericha, C. Bizon, M.D. Shattuck, and H.L. Swinney, Phys. Rev. Lett. {\bf 88}, 014302 (2002).
\bibitem{lohse} D. Lohse {\it et al.}, Phys. Rev. Lett. {\bf 93}, 198003 (2004).
\bibitem{royer} J. Royer {\it et al.}, Nature Phys. {\bf 1}, 164 (2005).
\bibitem{RHICtheory} P. Huovinen and P.V. Ruuskanen, Annu. Rev. Nucl. Part. Sci. {\bf 56}, 163 (2006).





\end{thebibliography}
\end{document}